\documentclass[11pt]{article}
\usepackage{hyperref}
\usepackage{amssymb}
\usepackage{graphicx}
\usepackage{float}
\usepackage{slashed}
\usepackage{amsmath}
\usepackage{tabularx}
\usepackage{breqn}
\usepackage{authblk}
\usepackage{array}
\usepackage{verbatim}
\usepackage{dsfont}
\usepackage{cite}
\usepackage{physics}
\usepackage[section]{placeins}
\usepackage[a4paper, total={6.8in, 10in}]{geometry}
\numberwithin{equation}{section}
\newcommand{\be}{\begin{equation}}
\newcommand{\ee}{\end{equation}}
\newcommand{\bea}{\begin{eqnarray}}
\newcommand{\eea}{\end{eqnarray}}
\newcommand{\ba}{\begin{aligned}}
\newcommand{\ea}{\end{aligned}}

\begin{document}
\title{Centripetal force on Casimir energies in $\kappa$-deformed rotating frame}

\author[1]{E. Harikumar\thanks{eharikumar@uohyd.ac.in}} 
\affil[1]{School of Physics, University of Hyderabad, Central University P.O. Hyderabad-500046, Telangana, India}
\author[2]{K. V. Shajesh\thanks{kvshajesh@gmail.com}}
\affil[2]{School of Physics and Applied Physics, Southern Illinois University-Carbondale, Carbondale, Illinois 62901, USA}
\author[1]{Suman Kumar Panja\thanks{sumanpanja19@gmail.com}} 
%\affil[1]{School of Physics, University of Hyderabad, \\Central University P.O, Hyderabad-500046, Telangana, India}

\maketitle
\begin{abstract}
We investigate the implications of a fundamental length scale on the centripetal force on a rotating Casimir apparatus in $\kappa$-space-time. We model the Casimir apparatus rotating with constant angular speed using appropriate $\kappa$-deformed coordinates. We find the $\kappa$-deformed centripetal force on a single plate, as well as for parallel plates. We show that the Casimir energy, including the divergent part (self energies of the plates) experiences centripetal forces like a conventional mass. We also find centripetal force on oriented parallel plates rotating with constant angular speed in $\kappa$-space-time. Results show that the mass-energy equivalence principle holds in the $\kappa$-space-time.
\end{abstract}

\section{Introduction} \label{intro}

Studying physics at the Planck length scale is one of the most intriguing topics.
At the Planck length, the four fundamental forces of nature are expected to unify. This unification program necessitates a consistent quantum theory of gravity. The development of a consistent quantum theory of gravity has been an active research topic. All approaches to quantum gravity have the inherent feature of exhibiting a fundamental length scale. Since models built on non-commutative space-times introduce a length scale, the relevance of such space-times to quantum gravity is being investigated vigorously \cite{CH-4-Glikman,CH-4-connes, CH-4-dop}.

Investigating the effect of a minimal length scale, introduced through non-commutativity in the physical phenomenon, is of intrinsic interest. A well-studied physical phenomenon where the length scale plays a significant role is the Casimir effect\cite{CH-4-hbg, CH-4-Mil, CH-4-kam,CH-4-Plunien}. The Casimir effect has been investigated using scalar field theory in non-commutative space-times, such as Moyal space-time and $\kappa$-deformed space-time\cite{CH-4-casadio, CH-4-fosco, CH-4-pinto, CH-4-skp}. It is shown that $\kappa$-deformed space-time is associated with the low-energy limit of specific quantum gravity models \cite{CH-4-jerzy} and its coordinates adhere to the following commutation relations
\begin{equation}
 [\hat{\bar{x}}^i, \hat{\bar{x}}^j]=0,~~ [\hat{\bar{x}}^0,\hat{\bar{x}}^i]=i a \hat{\bar{x}}^i,~~ a=\frac{1}{\kappa},\label{Chap-4-intro2}
\end{equation} 
where $a$  is the deformation parameter with the dimension of length. In \cite{CH-4-pinto, CH-4-skp}, authors studied the Casimir effect for a massless scalar field using the Green's function method and obtained a bound on the non-commutative parameter \cite{CH-4-skp}. Corrections to the Casimir force and energy in Doplicher-Fredenhagen-Roberts space-time \cite{CH-4-dop} have been obtained in \cite{CH-4-skp2}. In \cite{CH-4-skp3}, the gravitational force on two parallel plates in $\kappa$-deformed Rindler space-time is obtained using Green's function method. It is shown that $\kappa$-deformed Casimir energy, including the self energies associated with the plates, fall like a conventional mass.

Recently, many studies have been reported discussing the gravitational force on the Casimir apparatus \cite{CH-4-shajesh4,CH-4-shajesh1,CH-4-shajesh2a,CH-4-shajesh2b,CH-4-shajesh3,CH-4-saharian}. Using weak gravitational field approximation, the gravitational force on the Casimir energy is derived for Fermi coordinates and isotropic Schwarzschild coordinates \cite{CH-4-shajesh1}. In \cite{CH-4-shajesh2a,CH-4-shajesh2b}, authors found the gravitational force on two parallel plates in Rindler space-time using Green's function method. All these studies show that the Casimir energy, including the self-energy of the plates, falls in a weak gravitational field in accordance with the equivalence principle. In \cite{CH-4-shajesh3}, the Casimir apparatus is considered to rotate with constant angular speed such that the plates remain parallel to the tangent of the circle of motion. The centripetal force on parallel plates (for both parallel plates without orientation and with orientation) is calculated, and it is shown that the Casimir energy with its divergent counterpart (self energies of the plates) experiences the centripetal force exactly like a conventional mass. In this work, we investigate how the existence of the fundamental length scale in $\kappa$-deformed non-commutative space-time will modify the analysis and result. This is important since incorporating a fundamental length scale into these theories led to modifying the principle of relativity. Our study provides insight into how the fundamental length scale associated with quantum gravity influences the study related to the Casimir effect of parallel plate configurations in deformed space-time.

In this study, we generalize our earlier investigation of the Casimir effect in the $\kappa$-deformed Rindler space-time \cite{CH-4-skp3} to the case where the plates are rotating with the constant angular speed in the $\kappa$-space-time. Here, we follow the procedure used in \cite{CH-4-shajesh3} to investigate the centripetal force acting on the Casimir apparatus (single plate and two parallel plates). First, we define the local Lorentz coordinates \cite{CH-4-moller} and curvilinear coordinates (observer's reference frame) in $\kappa$-space-time. We find the general force expression measured in the local Lorentz coordinates using this definition of coordinates. Next, we calculate the centripetal force acting on a rotating single plate and then on the rotating two parallel plates using Green's function method. For this, we first find the reduced Green function by solving the equation of motion for the system containing a single plate in $\kappa$-deformed curvilinear coordinates. Using these Green's functions, we find the vacuum expectation value of $\kappa$-deformed energy-momentum tensor, which we use to find the centripetal force on the single plate. By repeating the same process for two parallel plates, we find the $\kappa$-deformed centripetal force on the two parallel plates rotating with constant angular speed. We show that divergent parts of the Casimir energy, coming from the self-energy of the plates, and the finite part of the Casimir energy experience the centripetal force like a conventional mass even in the $\kappa$-deformed space-time. We have also calculated the $\kappa$-deformed centripetal force on the rotating parallel plates which are oriented with an arbitrary angle to the tangent of the circle of rotation.

The organization of this paper is as follows. The next Section briefly reviews $\kappa$-deformed space-time and its symmetry algebra. In this Section, we also obtain $\kappa$-deformed line element for Minkowski space-time. By defining coordinate systems (local Lorentz coordinates and curvilinear coordinates) we construct $\kappa$-deformed force expression measured in the local Lorentz coordinates in Section \ref{kforce}. In Section \ref{rotcasimir}, we study the rotation of the Casimir apparatus and find the $\kappa$-deformed Lagrangian and symmetrized energy-momentum tensor in curvilinear coordinates. We consider modifications valid up to the first non-vanishing terms in the deformation parameter $a$. In Section \ref{centforce}, we first find the vacuum expectation value of the energy-momentum tensor in terms of Green's function, which we obtain by perturbatively solving the equation of motion for the case of a rotating single plate. We find the centripetal force on a rotating single plate using this in energy-momentum tensor expression. Then, we repeat the process and find the $\kappa$-deformed centripetal force on the parallel plates rotating with constant angular speed. Next, we consider the rotation of the oriented Casimir apparatus and find the centripetal force on the Casimir apparatus in Section \ref{oricentforce}. Our concluding remarks are given in Section \ref{conclu}. In \ref{append-A}, we derive the $\kappa$-deformed metric. We choose a specific realization for $\kappa$-deformed space-time. This allows us to write the non-commutative variables in terms of commutative variables, their derivatives, and the deformation parameter $a$. Using this in the generalized commutation relation for the phase space coordinates, the $\kappa$-deformed metric is constructed.

\section{$\kappa$-deformed space-time, symmetries, and dispersion relation} \label{symm}

In this Section, we summarise the important findings regarding the realization of $\kappa$-deformed space-time coordinates and the associated symmetry algebra \cite{CH-4-hopf,CH-4-Kova}. $\kappa$-deformed space-time is a Lie-algebraic type non-commutative space-time, with coordinates satisfying Eq.(\ref{Chap-4-intro2}). $\kappa$-deformed field theoretical models were constructed using the star product formalism \cite{CH-4-dimitrijevic,CH-4-dasz}. This formalism substitutes the usual concept of the pointwise product between the coordinates (and their functions) with the star product. This star product is invariant with the $\kappa$-Poincare algebra \cite{CH-4-dimitrijevic,CH-4-dasz}. Another method for studying field theoretical models involves representing non-commutative coordinates as functions of commutative coordinates and their derivatives \cite{CH-4-hopf,CH-4-Kova,CH-4-mel2}. This approach has been shown to be equivalent to the star product formalism in Ref.\cite{CH-4-mel3}. In this work, we will use the realization method \cite{CH-4-hopf,CH-4-Kova,CH-4-mel2,CH-4-mel3} and thus the coordinates of $\kappa$-deformed space-times are represented in terms of commutative coordinates $\bar{x}_{\mu}$ and their derivatives ($\bar{\partial}_{\mu}$) as \cite{CH-4-hopf,CH-4-Kova}
\begin{equation}\label{Chap-4-CFK2}
\begin{split}
 \hat{\bar{x}}_0=&\bar{x}_0\psi(A)+ia\bar{x}_j\bar{\partial}_j\gamma(A),\\
 \hat{\bar{x}}_i=&\bar{x}_i\varphi(A),
\end{split}
\end{equation}
where $A=ia\bar{\partial}_0=ap^{0}$. Here $\psi(A)$, $\gamma(A)$ and $\varphi(A)$ satisfy conditions
\begin{equation}\label{Chap-4-CFK3}
 \psi(0)=1,~\varphi(0)=1.
\end{equation}   
Substituting Eq.(\ref{Chap-4-CFK2}) in Eq.(\ref{Chap-4-intro2}) we obtain
\begin{equation}\label{Chap-4-CFK4}
 \frac{\varphi'(A)}{\varphi(A)}\psi(A)=\gamma(A)-1,
\end{equation} 
where $\varphi^{\prime}=\frac{d\varphi}{dA}$. Two possible realisations of $\psi(A)$ are $\psi(A)=1$ and $\psi(A)=1+2A$ \cite{CH-4-hopf,CH-4-Kova}. Here onwards, we choose $\psi(A)=1$. 
Allowed choices of $\varphi$ such as $e^{-A}$, $e^{-\frac{A}{2}}$, $1$, and $\frac{A}{e^A-1}$ lead to various ordering choices\cite{CH-4-hopf,CH-4-Kova}. For this study, we use $\varphi=e^{-\frac{A}{2}}$. Using this choice of $\varphi$ in Eq.(\ref{Chap-4-CFK2}), we find
\be
\hat{\bar{x}}_0=\bar{x}_0~~\text{and}~~ \hat{\bar{x}}_i=\bar{x}_i e^{-\frac{A}{2}}. \label{Chap-4-CFK6a}
\ee 
Symmetry algebra of the $\kappa$-Minkowski space-time is the $\kappa$-Poincare algebra \cite{CH-4-majid,CH-4-luk1,CH-4-luk2}. Thus, the commutation relations of the Poincare algebra get modified. Alternatively, the symmetry algebra can be realized with the standard Poincare algebra but with modified generators. This algebra is known as the undeformed $\kappa$-Poincare algebra \cite{CH-4-hopf,CH-4-Kova}.

The Lorentz generators of the undeformed $\kappa$-Poincare algebra \cite{CH-4-hopf,CH-4-Kova} satisfy
\begin{equation}\label{Chap-4-CFK7}
 [M_{\mu\nu},M_{\lambda\rho}]=M_{\mu\rho}\eta_{\nu\lambda}-M_{\nu\rho}\eta_{\mu\lambda}-M_{\mu\lambda}\eta_{\nu\rho}+M_{\nu\lambda}\eta_{\mu\rho}.
\end{equation} 
By demanding the commutation relation between the Lorentz generators and the $\kappa$-deformed space-time coordinate to be linear in $M_{\mu\nu}$ and $\hat{\bar{x}}_{\mu}$, one finds
\begin{equation}\label{Chap-4-CFK8a}
 [M_{\mu\nu},\hat{\bar{x}}_{\lambda}]=\hat{\bar{x}}_{\mu}\eta_{\nu\lambda}-\hat{\bar{x}}_{\nu}\eta_{\mu\lambda}+ia(M_{0\mu}\eta_{\nu\lambda}-M_{0\nu}\eta_{\mu\lambda}),
\end{equation}
and then using Jacobi identities, one finds the explicit form of the Lorentz generators of the undeformed $\kappa$-Poincare algebra as
\begin{equation}\label{Chap-4-CFK8}
\begin{split}
 M_{ij}=& \bar{x}_i\bar{\partial}_j-\bar{x}_j\bar{\partial}_i ~,\\
 M_{i0}=& \bar{x}_i\bar{\partial}_0\varphi\frac{e^{2A}-1}{2A}-\bar{x}_0\bar{\partial}_i\frac{1}{\varphi}+ia\bar{x}_i\bar{\partial}_k^2\frac{1}{2\varphi}-ia\bar{x}_k\bar{\partial}_k\bar{\partial}_i\frac{\gamma}{\varphi}.
\end{split} 
\end{equation}
Note that under the undeformed $\kappa$-Poincare algebra, the usual partial derivative, that is, $\bar{\partial}_{\mu}=\frac{\partial}{\partial_{\bar{x}^{\mu}}}$, does not transform as a $4$-vector. This necessitates the introduction of a new  derivative referred to as Dirac derivative, $\bar{D}_{\mu}$ \cite{CH-4-hopf,CH-4-Kova}, which transforms as a $4$-vector under this algebra, that is,
\begin{equation}\label{Chap-4-CFK9}
\begin{split}  
 [M_{\mu\nu},\bar{D}_{\lambda}]=&\bar{D}_{\mu}\eta_{\nu\lambda}-\bar{D}_{\nu}\eta_{\mu\lambda},\\
 [\bar{D}_{\mu},\bar{D}_{\nu}]=&0,
\end{split}
\end{equation}
where the components of the Dirac derivative are given to be
\begin{equation}\label{Chap-4-CFK10}
\begin{split}
 \bar{D}_0=&\bar{\partial}_0\frac{\sinh A}{A}+ia\bar{\partial}_k^2\frac{e^{-A}}{2\varphi^2},\\
 \bar{D}_i=&\bar{\partial}_i\frac{e^{-A}}{\varphi},
\end{split}
\end{equation}
satisfying
\begin{equation}\label{Chap-4-CFK10a}
 [\bar{D}_{\mu},\hat{\bar{x}}_{\nu}]=\eta_{\mu\nu}(ia\bar{D}_0+\sqrt{1+a^2\bar{D}_{\alpha}\bar{D}^{\alpha}})+ia\eta_{\mu 0}\bar{D}_{\nu}.
\end{equation}
The quadratic Casimir invariant corresponding to the undeformed $\kappa$-Poincare algebra is
\begin{equation}\label{Chap-4-CFK11}
 \bar{D}_{\mu}\bar{D}^{\mu}=\Box\left(1+\frac{a^2}{4}\Box\right),
\end{equation} 
where $\Box$ represents the $\kappa$-deformed Laplacian,
\begin{equation}\label{Chap-4-CFK12}
 \Box=\bar{\partial}_k^2\frac{e^{-A}}{2\varphi^2}-\bar{\partial}_0^2\frac{2(1-\cosh A)}{A^2}.
\end{equation}
Thus $\bar{D}_\mu$ and $M_{\mu\nu}$ define undeformed $\kappa$-Poincare algebra and the corresponding quadratic Casimir invariant, $\bar{D}_\mu \bar{D}^\mu=0$ in the momentum space, which is nothing but the deformed dispersion relation
\begin{equation}
\frac{4}{a^2}\sinh^2(\frac{A}{2}) -p_ip_i \frac{e^{-A}}{\varphi^2(A)}-m^2c^2 +\frac{a^2}{4}\left[\frac{4}{a^2}\sinh^2(\frac{A}{2}) -p_ip_i \frac{e^{-A}}{\varphi^2(A)}\right]^2=0.\label{Chap-4-dispersion}
\end{equation}
Here $p_i$ are the components of the momentum canonically conjugate to the commutative coordinates. The explicit form of the line element in $\kappa$-deformed space-time is defined \cite{CH-4-zuhair1} as (see Appendix-A for details)
\begin{equation}\label{Chap-4-N11}
\begin{aligned}
d\hat{s}^2&=g_{00}(\hat{\bar{y}})d\bar{x}^0d\bar{x}^0+\Big(g_{0i}(\hat{\bar{y}})\big(1-ap^0\big)-\frac{a}{2}g_{im}(\hat{\bar{y}})p^m\Big)e^{-\frac{ap^{0}}{2}}d\bar{x}^0d\bar{x}^i\\&+g_{i0}(\hat{\bar{y}})e^{-ap^{0}}d\bar{x}^id\bar{x}^0+g_{ij}(\hat{\bar{y}})e^{-2ap^{0}}d\bar{x}^id\bar{x}^j.
\end{aligned}
\end{equation}
Thus from Eq.(\ref{Chap-4-N6}) and Eq.(\ref{Chap-4-N11}), we find $g_{\mu\nu}(\hat{\bar{y}}^i)=g_{\mu\nu}(\bar{x}^i)$. We observe that the metric components only have an explicit dependence on spatial coordinates.  

In this study, we investigate the falling of the rotating Casimir apparatus in the $\kappa$-space-time. We consider a Casimir apparatus rotating (with respect to $\bar{x}$ axis) with constant angular speed in Minkowski space-time such that the surface of the plates (i.e., $\bar{y}$-$\bar{z}$ plane) remain parallel to the tangent to the circle of rotation. The metric components corresponding to the Minkowksi space-time is $\eta_{\mu \nu}=(-1,+1,+1,+1)$. Since the cross terms in the Minkowski metric tensor are zero (i.e., $\eta_{0i}=0$), the $\kappa$-deformed line element in Eq.(\ref{Chap-4-N11}) for the Minkowski space-time becomes \footnote{Under rotations $\kappa$-deformed space-time remains invariant. For calculational simplification, we employ realization (see Eq.(\ref{Chap-4-N6})) for non-commutative coordinate $\hat{y}_{\mu}$ in deriving the $\kappa$-deformed metric. This realization leads to non-zero off-diagonal terms in the $\kappa$-deformed metric. To retain rotational symmetry, we set off-diagonal components in the $\kappa$-deformed metric to zero.}
\begin{equation}\label{Chap-4-N12}
 d\hat{s}^2=g_{0 0}(\hat{\bar{y}})d\bar{x}^0d\bar{x}^0+g_{i j}(\hat{\bar{y}})e^{-2ap^0}d\bar{x}^id\bar{x}^j. 
\end{equation}
Using above equation we find the line element in the $\kappa$-deformed Minkowski space-time as
\begin{equation} \label{Chap-4-N13}
 d\hat{s}^2= -d\bar{t}^{2}+(1-2ap^{0})d\bar{x}_{i}^{2},~~ \text{where}~i=1,2,3.
\end{equation}
Note that corrections to the metric are considered up to first order in deformation parameter $a$ only. This correction is also dependent on the energy $p^{0}$ of the observer. We observe here that in the commutative limit, i.e., $a \rightarrow 0$, the above equation reduces to the usual form of the Minkowski metric.

\section{$\kappa$-deformed Force in the local Lorentz coordinates} \label{kforce}

In this Section, we derive the generic form of the force expression in the $\kappa$-deformed local Lorentz coordinate system. We consider a particular set of coordinates 
$\hat{\bar{x}}^{\mu} \equiv (\hat{\bar{t}},\hat{\bar{x}},\hat{\bar{y}},\hat{\bar{z}})$
with the metric given in Eq.(\ref{Chap-4-N13}). These coordinates are called the local Lorentz coordinates \cite{CH-4-moller} in $\kappa$-deformed space-time. Using Eq.(\ref{Chap-4-CFK6a}) we find the local Lorentz cordinates as $\hat{\bar{x}}^{\mu} \equiv \Big(\bar{t},\bar{x}e^{-\frac{ap^{0}}{2}},\bar{y}e^{-\frac{ap^{0}}{2}},\bar{z}e^{-\frac{ap^{0}}{2}}\Big)$ in $\kappa$-space-time. We also consider the $\kappa$-deformed curvilinear coordinates $\hat{x}^\mu \equiv (\hat{t},\hat{x},\hat{y},\hat{z})\equiv \Big(t,xe^{-\frac{ap^{0}}{2}},ye^{-\frac{ap^{0}}{2}},ze^{-\frac{ap^{0}}{2}}\Big)$ with the associated metric $\hat{g}_{\mu\nu}(x^{\alpha},a)$. Here $a$ is the deformation parameter in $\kappa$-space-time and $p^{0}$ energy of the observer/particle for probing the background space-time.

Additionally, we consider a transformation $\hat{\bar{x}}^{\mu}=\hat{\bar{x}}^{\mu}(\hat{x})$. We choose a point of reference \cite{CH-4-moller} i.e., any spatial point $\hat{{\bf x}}$  in the $\kappa$-defromed curvilinear coordinates. To obtain the force expression first we find the velocity of the point of reference $\hat{{\bf x}}$ as measured in the local Lorentz coordinates at time $\bar{t}$. For this using Eq.(\ref{Chap-4-CFK6a}) we find $\Delta \hat{\bar{\textbf{x}}}=e^{-\frac{ap^{0}}{2}}\Delta \bar{\textbf{x}}=e^{-\frac{ap^{0}}{2}}\Big(\Delta {x^i} \frac{\partial {\bf \bar{x}}}{\partial x^i} 
+ \Delta t \frac{\partial {\bf \bar{x}}}{\partial t}\Big)$ and $\Delta \hat{\bar{t}}=\Delta \bar{t}=\Delta {x^i} \frac{\partial \bar{t}}{\partial x^i}
+ \Delta t \frac{\partial \bar{t}}{\partial t}$. Thus, using the expression of $\Delta \hat{\bar{\textbf{x}}}$ and $\Delta \hat{\bar{t}}$ with $\Delta {x^i} =0$ for the point of reference, we find the velocity of the point of reference measured in the local Lorentz coordinates as
\be
\hat{\textbf{v}}(\bar{\textbf{x}}, \bar{t},a) 
= e^{-\frac{ap^{0}}{2}}\left. \frac{\Delta {\bf \bar{x}}}{\Delta \bar{t}} \right|_{\bf x}
= e^{-\frac{ap^{0}}{2}}\left[ \frac{\partial \bar{t}(x)}{\partial t} \right]^{-1}
   \frac{\partial {\bf \bar{x}}(x)}{\partial t}.
\label{Chap-4-def-vel}
\ee
The momentum for a small volume at 
a point of reference $\textbf{x}$ as measured in the $\kappa$-local 
Lorentz coordinates at time $\bar{t}$ is
\be
\hat{\bar{P}}^\alpha(\bar{x},a) 
= d\Sigma_\beta (\bar{x},a) \, \hat{t}^{\alpha \beta} (\bar{x},a)
= d\Sigma_\nu (x,a) \, \hat{t}^{\mu\nu} (x,a) \, 
  \frac{\partial \bar{x}^{\alpha}(x)}{\partial x^\mu}
= \hat{P}^\mu(x,a) \frac{\partial \bar{x}^{\alpha}(x)}{\partial x^\mu},\label{Chap-4-N13a}
\ee
where we have used tensor transformation properties. Here the spatial volume elements are constructed using the antisymmetric product of three space-like vectors, i.e., $d\Sigma_{\alpha}(\bar{x},a) =\sqrt{-\hat{\eta}(a)} \epsilon_{\alpha \beta \lambda \sigma}\,\delta_1 \bar{x}^{\beta} \,\delta_2 \bar{x}^{\lambda} \, \delta_3 \bar{x}^{\sigma}$,
and
$d\Sigma_\mu(x)=\sqrt{- g(x)} \,\epsilon_{\mu\nu\alpha\beta} 
   \,\delta_1 x^\nu \,\delta_2 x^\alpha \, \delta_3 x^\beta$,
where $\sqrt{-\hat{\eta}(a)}=(1-3ap^{0})$ obtained from Eq.(\ref{Chap-4-N13}) and $\hat{g}(x^{\alpha},a) = \mbox{det} \,\hat{g}_{\mu\nu}(x^{\alpha},a)$. Next we find the change in the momentum as
\be
\Delta \hat{\bar{P}}^{\alpha}(\bar{x},a)
= \Delta \bar{x}^i \frac{\partial}{\partial \bar{x}^i} \hat{\bar{P}}^{\alpha}(\bar{x},a)+\Delta \bar{t} \frac{\partial}{\partial \bar{t}} \hat{\bar{P}}^{\alpha}(\bar{x},a). \label{Chap-4-N14}
\ee
Thus, using the above equation, we find $\kappa$-deformed force density as measured by the observer at the point of reference in the local Lorentz coordinates at time $\bar{t}$ as
\bea
\hat{\bar{F}}^{\alpha}(\bar{x},a) = 
\frac{\Delta \hat{\bar{P}}^{\alpha}}{\Delta \bar{t}}\bigg|_{\bf x}
&=& 
\frac{\Delta \bar{x}^i}{\Delta \bar{t}}\bigg|_{\bf x}
\frac{\partial}{\partial \bar{x}^i} \hat{\bar{P}}^{\alpha}(\bar{x},a)\bigg|_{\bf x}
+ \frac{\partial}{\partial \bar{t}} \hat{\bar{P}}^{\alpha}(\bar{x},a)\bigg|_{\bf x}\nonumber \\
&=& \left[ \frac{\partial \bar{t}(x)}{\partial t} \right]^{-1}
\left[ \frac{\partial \bar{x}^i}{\partial t} 
       \frac{\partial}{\partial \bar{x}^i} \hat{\bar{P}}^{\alpha}(\bar{x},a)
       + \frac{\partial \bar{t}}{\partial t}
         \frac{\partial}{\partial \bar{t}} \hat{\bar{P}}^{\alpha}(\bar{x},a) \right]_{\bf x}\nonumber \\
&=& \left[ \frac{\partial \bar{t}(x)}{\partial t} \right]^{-1}
\left[ \frac{\partial}{\partial t} \hat{\bar{P}}^{\alpha}(\bar{x},a) \right]_{\bf x}\label{Chap-4-N15}
\eea
Using Eq.(\ref{Chap-4-N13a}) we rewrite the above equation as
\be
\hat{\bar{F}}^{\alpha}(\bar{x},a)=\left[ \frac{\partial \bar{t}(x)}{\partial t} \right]^{-1}
    \frac{\partial}{\partial t} 
    \left[ \hat{P}^\mu(x,a) \frac{\partial \bar{x}^{\alpha}(x)}{\partial x^\mu} \right],
\label{Chap-4-def-force}
\ee
To obtain the total force at time $\bar{t}$ in the $\kappa$-deformed local Lorentz coordinates, we integrate the force density over a surface $S$ described by constant $\bar{t}$ and find
\begin{eqnarray}
\hat{\bar{F}}^{\alpha}(\bar{t},a) 
&=& \int_S \left[ \frac{\partial \bar{t}(x)}{\partial t} \right]^{-1}
    \frac{\partial}{\partial t}
    \left[ \hat{P}^\mu(x,a) \frac{\partial \bar{x}^{\alpha}(x)}{\partial x^\mu} \right]
\nonumber \\
&=& \int d^3x 
\left[ \frac{\partial \bar{t}(x)}{\partial t} \right]^{-1}
\,\frac{\partial}{\partial t}
\left[ \sqrt{-\hat{g}(x,a)} \,\hat{t}^{\mu 0}(x,a) 
       \,\frac{\partial\bar{x}^a(x)}{\partial x^\mu}\right],
\label{Chap-4-FA}
\end{eqnarray}
where $\hat{t}^{\mu 0}(x,a)$ is the component of the $\kappa$-deformed energy-momentum tensor $\hat{t}^{\mu \nu}(x,a)$. For the case when both the metric tensor and the 
energy-momentum tensor are explicitly independent time coordinate $t$, we can reexpress the above equation as
\begin{eqnarray}
\hat{\bar{F}}^{\alpha}(\bar{t},a)
&=& \int d^3x \sqrt{-\hat{g}(x,a)} \,\hat{t}^{\mu 0}(x,a)        
\, \left[ \frac{\partial}{\partial t}
\,\frac{\partial\bar{x}^{\alpha}(x)}{\partial x^\mu} \right]
\left[ \frac{\partial \bar{t}(x)}{\partial t} \right]^{-1}.
\label{Chap-4-FA-ct}
\end{eqnarray}
Note that in the above expression of the force, corrections come due to non-commutativity from the determinant of the metric in the curvilinear coordinates, i.e., $\hat{g}(x,a)$ and from the energy-momentum tensor, i.e., $\hat{t}^{\mu 0}(x,a)$. In the limit, $a \rightarrow 0$, the above expression of the force reduces to the result obtained in \cite{CH-4-shajesh3}.

\section{Rotation of Casimir apparatus in $\kappa$-deformed space-time} \label{rotcasimir}

In this Section, we construct the Lagrangian describing the scalar field in the presence of the Casimir apparatus in $\kappa$-deformed local Lorentz coordinates.
Using this Lagrangian, we find the Lagrangian in $\kappa$-deformed curvilinear coordinates (reference frame of the observer positioned at the point of reference at the midpoint between two parallel plates) by considering the rotation of the Casimir apparatus. We find the corresponding $\kappa$-deformed energy-momentum tensor in $\kappa$-curvilinear coordinates. 

Using the expression of the quadratic Casimir invariant of the undeformed $\kappa$-Poincare algebra given in Eq.(\ref{Chap-4-CFK11}) we find the generalized massless Klein-Gordon equation in the $\kappa$-deformed local Lorentz coordinate system to be
\be
 \Box\left(1+\frac{a^2}{4}\Box\right)\bar{\phi}=0, \label{Chap-4-L1}
\ee
where $\Box$ is given in Eq.(\ref{Chap-4-CFK12}) and $\bar{\phi}$ is field defined in the local Lorentz coordinates. From the above equation, by considering the realization $\varphi=e^{-\frac{A}{2}}$, we find the equation of motion valid up to the first-nonvanishing order in deformation parameter $a$ as
\be
 \Big(\bar{\partial}_{\mu}\bar{\partial}^{\mu}(1+\frac{a^{2}}{4}\bar{\partial}_{\alpha}\bar{\partial}^{\alpha})+\frac{a^{2}}{12}\bar{\partial}_{0}^{4}\Big)\bar{\phi}=0. \label{Chap-4-L2}
\ee
The above equation of motion can be obtained from the following Lagrangian
\be
\mathcal{L}=-\frac{1}{2}\bar{\partial}_{\mu}\bar{\phi}\,\bar{\partial}^{\mu}\bar{\phi}+\frac{a^{2}}{8}\bar{\partial}_{\mu}\bar{\partial}^{\mu}\bar{\phi}\,\bar{\partial}_{\alpha}\bar{\partial}^{\alpha}\bar{\phi}+\frac{a^{2}}{24}\bar{\partial}_{0}^{2}\bar{\phi}\,\bar{\partial}_{0}^{2}\bar{\phi},  \label{Chap-4-L3}
\ee 
which we re-write as
\be
\mathcal{L}= -\frac{1}{2}\eta^{\mu\nu}\bar{\partial}_{\mu}\bar{\phi}\bar{\partial}_{\nu}\bar{\phi}+\frac{a^{2}}{8}\eta^{\mu\nu}\bar{\partial}_{\mu}\bar{\partial}_{\nu}\bar{\phi}\eta^{\alpha\beta}\bar{\partial}_{\alpha}\bar{\partial}_{\beta}\bar{\phi}+\frac{a^{2}}{24}\delta_{\mu 0}\delta_{\nu 0}\bar{\partial}^{\mu}\bar{\partial}^{\nu}\bar{\phi} \delta_{\alpha 0}\delta_{\beta 0}\bar{\partial}^{\alpha}\bar{\partial}^{\beta}\bar{\phi}. \label{Chap-4-L4}
\ee
Replacing $\eta^{\mu\nu}$ in the above equation with $\hat{\eta}^{\mu\nu} = \eta^{\mu \nu}+2ap^{0}(0,1,1,1)$ obtained from Eq.(\ref{Chap-4-N13}) we find the Lagrangian for scalar theory valid up to the first order in deformation parameter $a$ in the $\kappa$-deformed local Lorentz coordinate system as
\be
 \mathcal{\bar{L}}^{0}=\frac{1}{2}(\partial_{\bar{t}}\bar{\phi})^{2}-\frac{1}{2}\Big((\partial_{\bar{x}}\bar{\phi})^{2}+(\partial_{\bar{y}}\bar{\phi})^{2}+(\partial_{\bar{z}}\bar{\phi})^{2}\Big)-ap^{0}\Big((\partial_{\bar{x}}\bar{\phi})^{2}+(\partial_{\bar{y}}\bar{\phi})^{2}+(\partial_{\bar{z}}\bar{\phi})^{2}\Big). \label{Chap-4-L5}
\ee
To study the Casimir effect with two parallel plates, one needs to introduce these plates through their interactions with fields. The interactions of the plates for parallel plates are described by the interaction Lagrangian given by
\be
\mathcal{\bar{L}}_{int}=-\frac{1}{2}\bar{V}(\bar{x})\bar{\phi}(\bar{x})^{2},  \label{Chap-4-L6}
\ee
where parallel plates are modeled with the delta functions and thus, we have
\bea
\bar{V}(\bar{x})=\lambda_{1}\delta(\bar{z}-r_{1})+\lambda_{2}\delta(\bar{z}-r_{2}).\label{Chap-4-L7}
\eea
Here $\lambda_{1}$ and $\lambda_{2}$ are coupling constants with dimension of $\text{length}^{-1}$. Note that at initial time ($\bar{t}=0$) parallel plates are kept at $\bar{z}=r_{1}=r_{0}-\frac{L}{2}$ and $\bar{z}=r_{2}=r_{0}+\frac{L}{2}$. Here $r_{0}$, the midpoint between the two parallel plates, is the distance of the point of reference from the origin in local Lorentz coordinates, and $L$ is the distance between the two parallel plates. 

\begin{figure}[h!]
\begin{center}
\includegraphics[scale=.70]{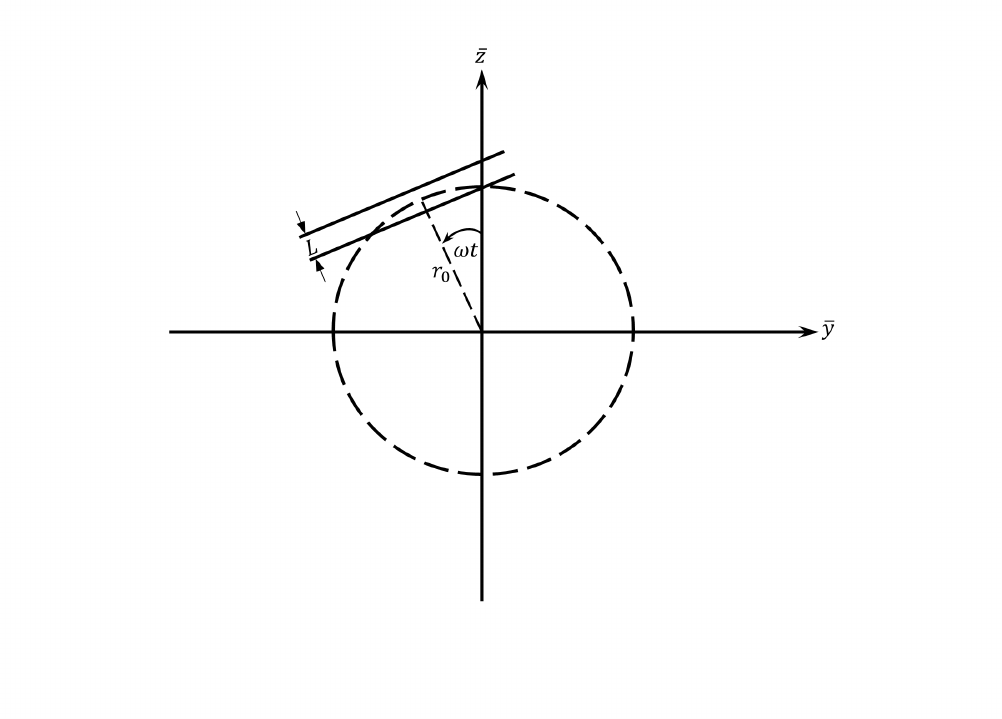}
\caption{\label{rot-c}
Casimir apparatus rotating with constant angular speed
$\omega$ about the $\bar{x}$ axis.}
\end{center}
\end{figure}

Next we consider that the Casimir apparatus rotating with constant angular speed $\omega$ about the $\bar{x}$ axis in the anti-clockwise direction, as indicated in Fig-\ref{rot-c}.
In Fig-\ref{rot-c}, the plates are kept perpendicular to the radial direction. Thus, the background potential is described as 
\begin{eqnarray}
\bar{V}(\bar{x}) 
&=& \lambda_{1}\,\delta (\bar{z}\cos\omega\bar{t}-\bar{y}\sin\omega\bar{t} - r_{1})
+\lambda_{2}\,\delta (\bar{z}\cos\omega\bar{t} - \bar{y}\sin\omega\bar{t} - r_{2})
\nonumber \\
&=& \lambda_{1}\,\delta \left(\bar{z}\cos\omega\bar{t}
      - \bar{y}\sin\omega\bar{t} - r_0 + \frac{L}{2} \right)
+\lambda_{2}\,\delta \left(\bar{z}\cos\omega\bar{t} 
      - \bar{y}\sin\omega\bar{t} - r_0 - \frac{L}{2} \right).
\label{Chap-4-V-bar}
\end{eqnarray} 
Thus, using Eq.(\ref{Chap-4-L5}), Eq.(\ref{Chap-4-L6}) and Eq.(\ref{Chap-4-V-bar}) we find complete Lagrangian in the $\kappa$-deformed space-time as
\be
\mathcal{\bar{L}}=\mathcal{\bar{L}}^{0}+\mathcal{\bar{L}}_{\text{int}}  \label{Chap-4-L8}
\ee
and we also find the corresponding action  in $\kappa$-deformed local Lorentz coordinates as
\bea
 \mathcal{S}_{\text{$\kappa$-LLC}}&=&\int d^{4}x \sqrt{-\hat{\eta}(a)}\,\mathcal{\bar{L}}(\bar{\phi},\bar{\partial}_{\mu}\bar{\phi},a) \nonumber \\
&=& \int d^4\bar{x} \sqrt{-\hat{\eta}(a)}
\left[
- \frac{1}{2} \hat{\eta}^{\alpha \beta}(a)\, 
\bar{\partial}_{\alpha} \bar{\phi}(\bar{x}) \bar{\partial}_{\beta} \bar{\phi}(\bar{x})
- \frac{1}{2} \bar{V}(\bar{x})\, \bar{\phi}(\bar{x})^2
\right], \label{Chap-4-L9}
\eea
where $\sqrt{-\hat{\eta}(a)}=(1-3ap^{0})$ valid upto first order in $a$ and $\bar{V}(\bar{x})$ is given in Eq.(\ref{Chap-4-V-bar}). In the limit $a \rightarrow 0$, we get back the action for commutative case \cite{CH-4-shajesh3}.

To obtain the Lagrangian and corresponding energy-momentum tensor in $\kappa$-deformed curvilinear coordinates (i.e., in the observer's reference frame, where the plates are at rest) we make the following transformations between the local Lorentz coordinates and curvilinear coordinates
\bea
y &=& +\bar{y}\cos\omega\bar{t}+\bar{z}\sin\omega\bar{t},~~~~x=\bar{x},
\label{Chap-4-trans-a}
\nonumber \\
z &=& -\bar{y}\sin\omega\bar{t}+\bar{z}\cos\omega\bar{t} - r_0,~~t=\bar{t}.
\label{Chap-4-trans-b}
\eea
Using the above equations we find the inverse transformations as
\bea
\bar{y} &=& y\cos\omega t - (z+r_0)\sin\omega t,~~~\bar{x}=x,
\label{Chap-4-in-trans-a}
\nonumber \\
\bar{z} &=& y\sin\omega t + (z+r_0)\cos\omega t,~~~\bar{t}=t,
\label{Chap-4-in-trans-b}
\eea
Using the above mentioned transformation in the action given in Eq.(\ref{Chap-4-L9}) and $\bar{\phi}=\phi$ for the scalar field, we find
\be
S_{\text{$\kappa$-Curvilinear}} = \int d^4x \sqrt{-\hat{g}(x,a)} \, {\cal{L}}(\phi(x),\partial_\mu \phi(x),a),\label{Chap-4-L10}
\ee
where the Lagrangian is
\be
{\cal{L}}(\phi(x)) 
= - \frac{1}{2} \hat{g}^{\mu\nu}(x,a) \partial_\mu \phi(x) \partial_\nu \phi(x)
- \frac{1}{2} V(x) \phi(x)^2, 
\label{Chap-4-L}
\ee
in which the explicit time-independent background potential is
\begin{equation}
V(x) = \lambda_1\,\delta \left(z + \frac{L}{2}\right) 
       + \lambda_2\,\delta \left(z - \frac{L}{2} \right),
\label{Chap-4-pot}
\end{equation}
In Eq.(\ref{Chap-4-L}) the expression of the metric tensor in $\kappa$-deformed curvilinear coordinates is
\be
\hat{g}_{\mu\nu}(x^{\alpha},a) = 
\left[ \begin{array}{cccc}
-\{1-(1-2ap^{0})\omega^2r^2\} &0& -(1-2ap^{0})\omega (z+r_0) & (1-2ap^{0})\omega y \\ 0&(1-2ap^{0})&0&0 \\
-(1-2ap^{0})\omega (z+r_0) &0& (1-2ap^{0}) & 0 \\ (1-2ap^{0})\omega y &0& 0 & (1-2ap^{0})
\end{array} \right],
\label{Chap-4-rot-met}
\ee
and the inverse of the above mentioned metric is
\be
\hat{g}^{\mu\nu}(x^{\alpha},a) =
\left[ \begin{array}{cccc}
-1 &0& -\omega (z+r_0) & \omega y \\ 
   0&(1+2ap^{0})&0&0 \\
-\omega (z+r_0) &0& \{(1+2ap^{0}) - \omega^2 (z+r_0)^2\} & \omega^2 y (z+r_0) \\ 
\omega y &0& \omega^2 y (z+r_0) & \{(1+2ap^{0}) - \omega^2 y^2\}
\end{array} \right],\label{Chap-4-rot-inmet}
\ee
where $r^2 = y^2 + (z+r_0)^2$, and $\sqrt{-\hat{g}(a)}=(1-3ap^{0})$.
Using Eq.(\ref{Chap-4-rot-met}) and Eq.(\ref{Chap-4-rot-inmet}) we find the nonzero components of the Christoffel symbols,
$\Gamma^2_{00}=-\omega^2 y, \Gamma^3_{00} = -\omega^2 (z+r_{0})$,
and $ \Gamma^3_{02} = - \Gamma^2_{03} = \omega$.

By varying the action given in Eq.(\ref{Chap-4-L10}) with respect of the background metric given in Eq.(\ref{Chap-4-rot-met}) we find the expression for the energy-momentum tensor as
\be
\hat{t}_{\mu\nu}=2\Big(-\frac{\partial \mathcal{L}}{\partial \hat{g}^{\mu\nu}(a)}+\frac{1}{2}\hat{g}_{\mu\nu}(a)\mathcal{L}\Big). \label{Chap-4-L11}
\ee
Next, using Eq.(\ref{Chap-4-L}) in the above equation, we find the expression for the energy-momentum tensor in the $\kappa$-deformed curvilinear coordinates to be
\be
\hat{t}_{\mu \nu}(x,a)
= \partial_{\mu} \phi(x) \partial_{\nu} \phi(x)
  + \hat{g}_{\mu \nu}(x,a) {\cal{L}}(\phi(x)).
\label{Chap-4-L13} 
\ee
where the Lagrangian $\mathcal{L}(\phi(x))$ is given in Eq.(\ref{Chap-4-L}). We have taken correction terms valid only up to the first order in $a$. In the commutative limit, i.e., $a \rightarrow 0$, we get back the commutative expression of energy-momentum tensor \cite{CH-4-shajesh3}.

\section{$\kappa$-deformed Centripetal force on Casimir apparatus} \label{centforce}

In this Section, we calculate the centripetal force acting on the Casimir apparatus rotating in the anti-clockwise direction with angular speed $\omega$ about the $\bar{x}$ axis. We derive the centripetal force on a single plate, and then we study the force experienced by the Casimir energy for the rotating Casimir apparatus. We note that the potential in terms of the curvilinear coordinates given in Eq.(\ref{Chap-4-pot}) is independent of time $t$. This explicit time independence of the potential given in Eq.(\ref{Chap-4-pot}) leads to the time-independent energy-momentum tensor. Thus, using the value of $\sqrt{-\hat{g}(a)}=(1-3ap^{0})$ and Eq.(\ref{Chap-4-trans-b}) in Eq.(\ref{Chap-4-FA-ct}) we obtain the expression of the force on the vacuum energy to be  
\begin{equation}
\hat{\bar{F}}^{\alpha}(\bar{t}) = (1-3ap^{0})\int d^3x \, \hat{t}^{\mu0}(a) 
\,\frac{\partial}{\partial t} \frac{\partial \bar{x}^{\alpha}(x)}{\partial x^\mu}.
\label{Chap-4-Ftmunu}
\end{equation}
We consider angular displacement to be very small, i.e., $\omega r_0 \leq 1$, and are interested in the leading order contributions to the force in  $\omega r_0$. Thus, we consider the contribution only from the zeroth order term of the energy-momentum tensor. 

Using Eq.(\ref{Chap-4-in-trans-b}) in Eq.(\ref{Chap-4-Ftmunu}) we find the components of force to be $\hat{\bar{F}}^{0}(\bar{t})=0$, $\hat{\bar{F}}^{1}(\bar{t})=0$, and
\be
\hat{\bar{F}}^{2}(\bar{t})=(1-3ap^{0})\int d^3x \, \hat{t}^{(0)}_{00}(a) 
\,\bigg(\frac{\partial}{\partial t} \frac{\partial \bar{y}}{\partial t}\bigg) + \omega^{2}\mathcal{O}(\omega r_{0}),
\ee
and 
\be
\hat{\bar{F}}^{3}(\bar{t})=(1-3ap^{0})\int d^3x \, \hat{t}^{(0)}_{00}(a) 
\,\bigg(\frac{\partial}{\partial t} \frac{\partial \bar{z}}{\partial t}\bigg) + \omega^{2}\mathcal{O}(\omega r_{0}).
\ee
Next we use Eq.(\ref{Chap-4-in-trans-b}) in the above two equations and find the centripetal force on the vacuum energy to be 
\bea
\hat{\bar{\textbf{F}}}(\bar{t}) &=&(1-3ap^{0})\,\Big\{-\omega^{2}\int d^3x \, \hat{t}^{(0)}_{00}(a)\,(r_{0}+z)\,(\cos\omega t \hat{\bar{\textbf{z}}}-\sin\omega t \hat{\bar{\textbf{y}}})+\omega^{2} \mathcal{O}(\omega r_0)\Big\} \nonumber \\
&=& (1-3ap^{0})\Big\{- \omega^2 r_{0} \, \hat{\bar{\textbf{r}}}(\omega \bar{t}) \int d^3x \hat{t}_{00}^{(0)}(a)-\omega^2 \hat{\bar{\textbf{r}}}(\omega \bar{t}) \int d^3x z \hat{t}_{00}^{(0)}(a)+\omega^{2} \mathcal{O}(\omega r_0)\Big\}, \label{Chap-4-Ft00-1}
\eea
where the unit vector $\hat{\bar{\textbf{r}}}(\omega t) = \cos\omega t\,\hat{\bar{\textbf{z}}}-\sin\omega t\,\hat{\bar{\textbf{y}}}$. We define the total energy and the center of inertia measured in the curvilinear coordinates as 
\be
\hat{E}^{(0)}_{\text{tot}}(a) = (1-3ap^{0})\int d^3x \, \hat{t}_{00}^{(0)}(a)
\qquad {\rm and} \qquad
z_{\rm cm} = \frac{(1-3ap^{0})}{\hat{E}^{(0)}_{\text{tot}}(a)} \int d^3x \,z \, \hat{t}_{00}^{(0)}(a),
\label{Chap-4-E-Q}
\ee
respectively. Thus, using above equations in Eq.(\ref{Chap-4-Ft00-1}) we rewrite the centripetal force as
\be
\hat{\bar{\textbf{F}}}(\bar{t},a) =
- \omega^2 \,\left[ r_0 + z_{\text{cm}} \right] 
  \,\hat{\bar{\textbf{r}}}(\omega \bar{t}) \, \hat{E}_{\text{tot}}^{(0)}(a).
\ee 
Here in the above equation defining $r_0 + z_{\rm cm}=\bar{r}_{\rm cm}$, as the center of inertia in the local Lorentz coordinates, we find
\be 
\hat{\bar{\textbf{F}}}(\bar{t},a)= - \omega^2 \, \bar{r}_{\text{cm}} 
  \,\hat{\bar{\textbf{r}}}(\omega \bar{t}) \, \hat{E}_{\text{tot}}^{(0)}(a). 
\label{Chap-4-Ft00}
\ee
To obtain the centripetal force on the Casimir apparatus, we need to find $\hat{E}_{tot}^{(0)}(a)$, i.e., $\hat{t}_{00}^{(0)}(a)$, which is the energy density of the parallel 
plates. We start with the zeroth order Lagrangian (obtained from Eq.(\ref{Chap-4-L}))
\be
\hat{\mathcal{L}}^{(0)}(a)=\frac{1}{2}(\partial_{t}\phi)^{2}-\frac{1}{2}\Big((\partial_{x}\phi)^{2}+(\partial_{y}\phi)^{2}+(\partial_{z}\phi)^{2}\Big)-\frac{1}{2}V(x)\phi^{2}-ap^{0}\Big((\partial_{x}\phi)^{2}+(\partial_{y}\phi)^{2}+(\partial_{z}\phi)^{2}\Big), \label{Chap-4-L0th}
\ee
where $V(x)$ is given in Eq.(\ref{Chap-4-pot}). From the above equation we find the equation of motion as
\be
\bigg\{-\partial_{0}^{2}+\partial_{x}^{2}+\partial_{y}^{2}+\partial_{z}^{2} -V(x)+2ap^{0}\Big(\partial_{x}^{2}+\partial_{y}^{2}+\partial_{z}^{2}\Big)\bigg\} \phi =0. \label{Chap-4-othEOM}
\ee
Next using Eq.(\ref{Chap-4-L0th}) and Eq.(\ref{Chap-4-othEOM}) in Eq.(\ref{Chap-4-L13})
we calculate the $\hat{t}_{00}^{(0)}$ component as
\be
\hat{t}^{(0)}_{00}(a)=\frac{1}{2}\partial_{t}\phi \partial_{t}\phi -\frac{1}{2}\phi \partial_{t}^{2}\phi +\frac{1}{4}\partial_{i}^{2}(\phi^{2})+\frac{ap^{0}}{2}\partial_{i}^{2}(\phi^{2}).\label{Chap-4-othA15}
\ee
Using $\langle\phi(x)\phi(x^{\prime})\rangle=\frac{1}{i}\,G(x,x^{\prime})$, we find the vacuum expectation value of $\hat{t}^{(0)}_{00}(a)$ as 
\begin{multline}
\langle\hat{t}^{(0)}_{00}(a)\rangle=\frac{1}{2i}\Big\{(\partial_{t}\partial_{t^{\prime}}-\partial_{t}^{2})\hat{G}^{(0)}(x,x^{\prime},a)\Big\}\Big\vert_{x=x^{\prime}}+\frac{1}{4i}\partial_{i}^{2}\Big\{\hat{G}^{(0)}(x,x^{\prime},a)\Big\vert_{x=x^{\prime}}\Big\} + \frac{ap^{0}}{2i}\partial_{i}^{2}\Big\{\hat{G}^{(0)}(x,x^{\prime},a)\Big\vert_{x=x^{\prime}}\Big\}. \label{Chap-4-othA16}
\end{multline}
The Green's function ($\hat{G}^{(0)}(x,x^{\prime},a)$) corresponding to Eq.(\ref{Chap-4-othEOM}) satisfy
\begin{multline}
\bigg\{-\partial_{0}^{2}+\partial_{x}^{2}+\partial_{y}^{2}+\partial_{z}^{2} -\lambda_{1}\delta(z+\frac{L}{2})-\lambda_{2}\delta(z-\frac{L}{2})\\+2ap^{0}\Big(\partial_{x}^{2}+\partial_{y}^{2}+\partial_{z}^{2}\Big)\bigg\}\hat{G}^{(0)}(x,x^{\prime},a)=(1+3ap^{0})\delta(t-t^{\prime})\delta(z-z^{\prime})\delta(x_{\perp}-x_{\perp}^{\prime}), \label{Chap-4-othA7}
\end{multline}
where $x_{\perp}=x,y$. Note that correction terms on the right-hand side are coming from the inverse of $\sqrt{-\hat{g}^{(0)}}=(1-3ap^{0})$, where $\hat{g}^{(0)}=\text{Det}\,\hat{g}^{(0)}_{\mu \nu}$, which is independent of $\omega$. We observe that $V(x)=\lambda_{1}\delta(z+\frac{L}{2})+\lambda_{2}\delta(z-\frac{L}{2})$ has only $z$ dependency. Thus, the Fourier transformation of Green's function is 
\bea
\hat{G}^{(0)}(x,x^\prime , a)
= \int_{-\infty}^{+\infty}
  \frac{d\omega}{2 \pi} \int \frac{d^2 k_\perp}{(2 \pi)^2}
  \, e^{-i \omega (t - t^\prime)}
  e^{i {\bf k}_\perp \cdot ({\bf x}_\perp - {\bf x}_\perp^\prime)}
  \hat{g}^{(0)}(z,z^\prime , a), \label{Chap-4-othGreen}
\eea
where ${k}_\perp =\sqrt{k_{x}^{2}+k_{y}^{2}} $ and $\hat{g}^{(0)}(z,z^{\prime},a)$ is the reduced Green's function.
Next using Eq.(\ref{Chap-4-othA16}) and Eq.(\ref{Chap-4-othGreen}) in Eq.(\ref{Chap-4-E-Q}) we find the total energy per unit area as 
\be
\frac{\hat{E}^{(0)}_{\text{tot}}(a)}{A}=(1-3ap^{0})\int \frac{dz d \omega d^{2}k_{\perp}}{8 \pi^{3}}  \Bigg[\frac{1}{2i}\Big(2\omega^{2} \hat{g}^{(0)}(z,z,a)\Big)+\frac{1}{4i}\partial_{i}^{2}\Big(\hat{g}^{(0)}(z,z,a)\Big) +\frac{ap^{0}}{2i}\partial_{i}^{2}\Big(\hat{g}^{(0)}(z,z,a))\Big)\Bigg]. \label{Chap-4-0thA17}
\ee
Next, using the Gauss Divergence theorem with $\omega=i \zeta$ in the above equation we find the total energy per unit area
\be
\frac{\hat{E}^{(0)}_{\text{tot}}}{A}=-(1-3ap^{0})\Bigg[\int \frac{d \zeta d^{2}k_{\perp}}{16 \pi^{3}}2\zeta^{2}\int^{+\infty}_{-\infty}dz\,\hat{g}^{(0)}(z,z,a)\Bigg]. \label{Chap-4-A18}
\ee
Using Eq.(\ref{Chap-4-othA16}) and Eq.(\ref{Chap-4-othGreen}) in Eq.(\ref{Chap-4-E-Q}) we find the center of inertia (energy) measured in the curvilinear coordinates as
\be
z_{\text{cm}}=-\frac{(1-3ap^{0})}{\hat{E}^{(0)}_{\text{tot}}}\Bigg[\int \frac{d \zeta d^{2}k_{\perp}}{16 \pi^{3}}2\zeta^{2}\int^{+\infty}_{-\infty}z\,\hat{g}^{(0)}(z,z,a)\,dz\Bigg]. \label{Chap-4-inertiaA18}
\ee
We find the centripetal force acting on the Casimir apparatus using the above equation in Eq.(\ref{Chap-4-Ft00}). We observe that we need to find a modified expression of the total energy per unit area associated with plates to obtain the force expression. Modified total energy per unit area depends on the reduced Green's function solutions in different regions of the plates.

\subsection{Centripetal force on a single plate} \label{singlecentforce}

In this Subsection, we find the $\kappa$-deformed centripetal force on the single plate. We consider a single plate rotating with constant angular speed ($\omega$) described by a single delta function in the potential in Eq.(\ref{Chap-4-pot}). We consider $L \rightarrow 0$, $r_0=r_1$, and $\lambda_{2}=0$ i.e., from Eq.(\ref{Chap-4-pot}) we have potential $V(z) = \lambda_{1} \, \delta (z)$ for single plate at $z=0$ in curvilinear coordinates (frame of reference of the observer positioned at midpoint of the single plate). Thus for single plate using  Eq.(\ref{Chap-4-othA7}) and Eq.(\ref{Chap-4-othGreen}) we find that the reduced Green's function satisfy
\be
-\bigg\{\partial_{z}^{2}-\bar{k}^{2}-\lambda_{1}\delta(z)+ 2ap^{0}\Big(\partial_{z}^{2}-k_{\perp}^{2}\Big)\bigg\} \hat{g}^{(0)}(z,z^{\prime},a)=(1+3ap^{0})\delta(z-z^{\prime}), \label{Chap-4-0thA9}
\ee
where $\bar{k} =\sqrt{ {\bf k}_\perp^2 + \zeta^2}$ and $\zeta=-i\omega$. Note that by replacing Eq.(\ref{Chap-4-othGreen}) in Eq.(\ref{Chap-4-othA7}), the dependence on the other coordinates ($t$, $x$, and $y$) for Green's function is canceled out. We find that the reduced Green's function ($\hat{g}^{(0)}(z,z^{\prime},a)$) depends only on $z$. Thus, from Eq.(\ref{Chap-4-othA16}), we observe that the zeroth-order energy-momentum tensor is diagonal and a function of $z$ since the vacuum expectation value of $\hat{t}^{(0)}_{\alpha \beta}$ is a function of $z$ only. We solve the above equation  perturbatively assuming the solution $\hat{g}^{(0)}(z,z^{\prime},a)$ to be of the form
\be
\hat{g}^{(0)}(z,z^{\prime},a)=g^{(0)}_{0}(z,z^{\prime})+ap^{0}g^{(0)}_{1}(z,z^{\prime}). \label{Chap-4-L18}
\ee
Substituting Eq.(\ref{Chap-4-L18}) in Eq.(\ref{Chap-4-0thA9}), we obtain the following equations
\be
 -\bigg\{\partial_{z}^{2}-\bar{k}^{2}-\lambda_{1}\delta(z)\bigg\}g^{(0)}_{0}(z,z^{\prime})=\delta(z-z^{\prime}), \label{Chap-4-L19}
\ee
and
\be
 -\bigg[\Big(\partial_{z}^{2}-\bar{k}^{2}-\lambda_{1}\delta(z)\Big)g^{(0)}_{1}(z,z^{\prime})+2\Big\{(\partial_{z}^{2}-k_{\perp}^{2} \Big\} g^{(0)}_{0}(z,z^{\prime})\bigg]=3\,\delta(z-z^{\prime}). \label{Chap-4-L20}
\ee
We solve Eq.(\ref{Chap-4-L19}) and find the solution for  $\hat{g}^{(0)}_{0}(z,z^{\prime})$ in different regions for a single plate placed at $z=0$ in curvilinear coordinates as \cite{CH-4-shajesh3}
\bea
 g^{(0)}_{0}(z,z^{\prime}) &=& \frac{1}{2\bar{k}}e^{-\bar{k}|z-z^{\prime}|}-\frac{\lambda_{1}}{\lambda_{1}+2\bar{k}}\frac{1}{2 \bar{k}}e^{\bar{k}(z+z^{\prime})},~~~\text{where}~\left\lbrace z,z^{\prime}\right\rbrace < 0, \nonumber \\
 &=&\frac{1}{2\bar{k}}e^{-\bar{k}|z-z^{\prime}|}-\frac{\lambda_{1}}{\lambda_{1}+2\bar{k}}\frac{1}{2 \bar{k}}e^{-\bar{k}(z+z^{\prime})},~~\text{where}~\left\lbrace z,z^{\prime}\right\rbrace >0. \label{Chap-4-L29}
\eea
Using Eq.(\ref{Chap-4-L19}) and Eq.(\ref{Chap-4-L29}) in Eq.(\ref{Chap-4-L20}), we find expression of  $\hat{g}^{(0)}_{1}(z,z^{\prime})$ as
\be
g^{(0)}_{1}(z,z^{\prime})=g^{(0)}_{0}(z,z^{\prime})+2\int d\tilde{z}\,g^{(0)}_{0}(z,\tilde{z})\Big(\zeta^{2}+\lambda_{1} \delta(\tilde{z})\Big)g^{(0)}_{0}(\tilde{z},z^{\prime}). \label{Chap-4-L23}
\ee
Note $\tilde{z}$ takes value between $z$ and $z^{\prime}$. Integration operation in the second term of the above equation ranges from $z$ to $z^{\prime}$ when $z^{\prime} > z$ and it ranges $z^{\prime}$ to $z$ when $z > z^{\prime}$. Thus, we observe that the delta function $\big(\delta(\tilde{z})\big)$ dependent term inside the integration does not contribute.

Thus, using Eq.(\ref{Chap-4-L29}), and Eq.(\ref{Chap-4-L23}) in Eq.(\ref{Chap-4-L18}) we find the $\kappa$-deformed reduced Green's function, for Euler-Lagrange equation given in Eq.(\ref{Chap-4-0thA9}) valid upto first order in $a$ as
\bea
\hat{g}^{(0)}(z,z^{\prime},a)&=&(1+ap^{0})g^{(0)}_{0}(z,z^{\prime})+2ap^{0}\Bigg[\int d\tilde{z}g^{(0)}_{0}(z,\tilde{z})\Big(\zeta^{2}+\lambda_{1} \delta(\tilde{z})\Big)g^{(0)}_{0}(\tilde{z},z^{\prime})\Bigg]. \label{Chap-4-L24}
\eea
Using Eq.(\ref{Chap-4-L29}) and Eq.(\ref{Chap-4-L24}) in Eq.(\ref{Chap-4-A18}) we find 
\be
\frac{\hat{E}^{(0)}_{\text{tot}}}{A}=-(1-2ap^{0})\Bigg[\int \frac{d \zeta d^{2}k_{\perp}}{16 \pi^{3}}\,\frac{\zeta^{2}}{\bar{k}^{2}}\bigg\{\int^{+\infty}_{-\infty}\bar{k}\,dz-\frac{\lambda_{1}}{\lambda_{1}+2\bar{k}}\bigg\} \Bigg]  \label{Chap-4-othE1}
\ee
Using spherical polar coordinates for phase space we re-express $\int^{+ \infty}_{-\infty} \frac{d \zeta d^{2}k_{\perp}}{16 \pi^{3}}\frac{\zeta^{2}}{\bar{k}^{2}}$ as $\frac{1}{12\pi^{2}}\int^{\infty}_{0} \bar{k}^{2} d\bar{k}$ and find
\bea
\frac{\hat{E}^{(0)}_{\text{tot}}}{A}&=&-(1-2ap^{0})\Bigg[\frac{1}{12 \pi^{2}}\int^{\infty}_{0}\bar{k}^{3}d\bar{k} \int^{+\infty}_{-\infty}dz-\frac{1}{12 \pi^{2}}\int^{\infty}_{0}\bar{k}^{2}d\bar{k}\frac{\lambda_{1}}{\lambda_{1}+2\bar{k}} \Bigg] \nonumber \\
&=& \frac{\hat{E}^{(0)}_{\text{bulk}}}{A}+\frac{\hat{E}^{(0)}_{\text{d1}}}{A}\label{Chap-4-othE2}
\eea
Note here that the first term in the above equation is divergent, which we define as bulk energy per unit area  $\Big(\frac{\hat{E}^{(0)}_{\text{bulk}}}{A}\Big)$ of the system. In the above equation, self energy per unit area of the plate $\Big(\frac{\hat{E}^{(0)}_{\text{d1}}}{A}\Big)$ can be is reexpressed as
\bea
 \frac{\hat{E}^{(0)}_{\text{d1}}}{A}&=& (1-2ap^{0})\frac{\lambda_{1}}{12 \pi^{2}}\int^{\infty}_{0} \frac{\bar{k}^{2} d\bar{k}}{\lambda_{1}+2\bar{k}} \nonumber \\
 &=&(1-2ap^{0})\frac{\lambda_{1}}{96 \pi^{2}}\int^{\infty}_{0}\frac{Y^{2} dY}{\lambda_{1}+Y},\label{Chap-4-L60}
\eea
where $Y=2\bar{k} $. Note from the above equation, the energy per unit area associated with a single plate is a divergent quantity. Next using Eq.(\ref{Chap-4-L29}) and Eq.(\ref{Chap-4-L24}) in Eq.(\ref{Chap-4-inertiaA18}) we find $z_{\text{cm}}=0$. By substituting the physically relevant part $\frac{\hat{E}^{(0)}_{\text{d1}}}{A}$ of $\frac{\hat{E}^{(0)}_{\text{tot}}}{A}$, in Eq.(\ref{Chap-4-Ft00}), we find the $\kappa$-deformed centripetal force on the single plate as
\bea
\hat{\bar{\textbf{F}}}(\bar{t},a) &=& - \hat{\bar{\textbf{r}}}(\omega\bar{t}) 
       \, \omega^2 \bar{r}_{cm} \, \hat{E}_{\text{d1}}^{(0)}\nonumber \\
       &=&  -(1-2ap^{0})\,\hat{\bar{\textbf{r}}}(\omega\bar{t}) 
       \, \omega^2 \bar{r}_{\text{cm}} \,\bigg(\frac{\lambda_{1}}{96 \pi^{2}}\int^{\infty}_{0}\frac{Y^{2} dY}{\lambda_{1}+Y}\bigg),
\eea
where $\bar{r}_{\text{cm}}=r_0=r_{1}$. A negative sign in the force expression implies that the plate is experiencing the centripetal force towards the origin in local Lorentz coordinates. Note that we have taken modifications valid up to the first order in deformation parameter $a$. In the limit $a \rightarrow 0$, the above $\kappa$-deformed centripetal force on the single plate reduces to the commutative result \cite{CH-4-shajesh3}.

\subsection{Centripetal force on Parallel plates} \label{doublecentforce}

In this Subsection, we find the $\kappa$-deformed centripetal force on the parallel plates. The two parallel plates separated by a distance $L$ and rotating
with constant angular speed $\omega$ about the $\bar{x}$ axis is described by the potential in Eq.(\ref{Chap-4-pot}). To obtain the centripetal force on the plates, we find the reduced Green's functions in different regions. Thus, using Eq.(\ref{Chap-4-othGreen}) in Eq.(\ref{Chap-4-othA7}) we find that the reduced Green's function satisfy
\be
-\bigg\{\partial_{z}^{2}-\bar{k}^{2}-\lambda_{1}\delta(z+\frac{L}{2})-\lambda_{2}\delta(z-\frac{L}{2})+ 2ap^{0}\Big(\partial_{z}^{2}-k_{\perp}^{2}\Big)\bigg\} \hat{g}^{(0)}(z,z^{\prime},a)=(1+3ap^{0})\delta(z-z^{\prime}), \label{Chap-4-0thA9p}
\ee
where $\bar{k} =\sqrt{ {\bf k}_\perp^2 + \zeta^2}$. Next, by following the same procedure discussed in Subsection \ref{singlecentforce},  we perturbatively solve the above equation and find 
\bea
\hat{g}^{(0)}(z,z^{\prime},a)&=&(1+ap^{0})g^{(0)}_{0}(z,z^{\prime})+2ap^{0}\Bigg[\int d\tilde{z}~g^{(0)}_{0}(z,\tilde{z})\bigg\{\zeta^{2}+\lambda_{1}\delta\Big(z+\frac{L}{2}\Big)+\lambda_{2}\delta\Big(z-\frac{L}{2}\Big)\bigg\}g^{(0)}_{0}(\tilde{z},z^{\prime})\Bigg],\nonumber \\ \label{Chap-4-0thA10}
\eea
where  
\bea
g^{(0)}_{0}(z,z^{\prime})&=& \Bigg[\frac{1}{2\bar{k}}e^{-\bar{k}|z-z^{\prime}|}-\frac{1}{2\bar{k}\Delta}e^{\bar{k}(z+z^{\prime}+L)}\bigg\{\frac{\lambda_{1}}{2\bar{k}}\Big(1+\frac{\lambda_{2}}{2\bar{k}}\Big)e^{2\bar{k}L} + \frac{\lambda_{2}}{2\bar{k}}\Big(1-\frac{\lambda_{1}}{2\bar{k}}\Big)\bigg\}  \Bigg],~\text{when}~\left\lbrace z, z^{\prime}\right\rbrace <-\frac{L}{2}. \nonumber \\\label{Chap-4-0thA11}\\
&=& \Bigg[\frac{1}{2\bar{k}}e^{-\bar{k}|z-z^{\prime}|}-\frac{1}{2\bar{k}\Delta}\bigg\{\frac{\lambda_{1}}{2\bar{k}}\Big(1+\frac{\lambda_{2}}{2\bar{k}}\Big)e^{-\bar{k}(z+z^{\prime}-L)} + \frac{\lambda_{2}}{2\bar{k}}\Big(1+\frac{\lambda_{1}}{2\bar{k}}\Big)e^{\bar{k}(z+z^{\prime}+L)}\nonumber \\
&&~~~~ -\frac{\lambda_{1}\lambda_{2}}{4\bar{k}^{2}}2\cosh \bar{k}(z-z^{\prime})\bigg\}  \Bigg], ~~~~~~~~~~~~~~~~~~~~~~~~~~~~~~~~~~~~~~~~~~\text{when}~-\frac{L}{2}<\left\lbrace z, z^{\prime}\right\rbrace <\frac{L}{2}. \nonumber \\ \label{Chap-4-0thA12}\\
&=& \Bigg[\frac{1}{2\bar{k}}e^{-\bar{k}|z-z^{\prime}|}-\frac{1}{2\bar{k}\Delta}e^{-\bar{k}(z+z^{\prime}-L)}\bigg\{\frac{\lambda_{1}}{2\bar{k}}\Big(1-\frac{\lambda_{2}}{2\bar{k}}\Big) + \frac{\lambda_{2}}{2\bar{k}}\Big(1+\frac{\lambda_{1}}{2\bar{k}}\Big)e^{2\bar{k}L}\bigg\}  \Bigg],~\text{when}~\left\lbrace z, z^{\prime}\right\rbrace > \frac{L}{2}. \nonumber \\ \label{Chap-4-0thA13}
\eea
and
\be
\Delta=\Big(1+\frac{\lambda_{1}}{2\bar{k}}\Big)\Big(1+\frac{\lambda_{2}}{2\bar{k}}\Big)e^{2\bar{k}L}-\frac{\lambda_{1} \lambda_{2}}{4\bar{k}^{2}}. \label{Chap-4-0thA14} 
\ee
Note that as in the previous case (Eq.(\ref{Chap-4-L24}) for single plate) here also the integral in the second term in Eq.(\ref{Chap-4-0thA10}) has range either $z \rightarrow z^{\prime}$ for $z^{\prime}> z$ or $z^{\prime} \rightarrow z$ when $z > z^{\prime}$. Thus, only the first term will contribute in finding $\hat{g}^{(0)}(z,z, a)$ from the above expression. 

To obtain the force on the parallel plates we need to find the $\kappa$-deformed total energy ($\hat{E}^{(0)}_{\text{tot}}(a)$) given in Eq.(\ref{Chap-4-A18})) associated with the two parallel plates. Thus, using Eq.(\ref{Chap-4-0thA11}), Eq.(\ref{Chap-4-0thA12}) and Eq.(\ref{Chap-4-0thA13}) we first obtain 
 \be
 2 \zeta^{2} \int^{+\infty}_{-\infty}dz~g^{(0)}_{0}(z,z)=\frac{\zeta^{2}}{\bar{k}^{2}}\int^{+\infty}_{-\infty} \bar{k}dz -\frac{\zeta^{2}}{\bar{k}^{2}}\frac{1}{\Delta_{0}}\Bigg[\frac{\lambda_{1}}{2\bar{k}}+\frac{\lambda_{2}}{2\bar{k}}+2\frac{\lambda_{1}}{2\bar{k}}\,\frac{\lambda_{2}}{2\bar{k}}\bigg\{1-e^{-2\bar{k}L}\bigg\}-2\bar{k}L\,\frac{\lambda_{1}}{2\bar{k}}\frac{\lambda_{2}}{2\bar{k}}e^{-2\bar{k}L}\Bigg], \label{Chap-4-A19}
 \ee
where $\Delta_{0}=e^{-2\bar{k}L}\Delta$. We use Eq.(\ref{Chap-4-0thA10}) and Eq.(\ref{Chap-4-A19}) in Eq.(\ref{Chap-4-A18}) and then using spherical polar coordinate for the phase space (see discussion after Eq.(\ref{Chap-4-othE1})) find total energy per unit area as
\begin{multline}
 \frac{\hat{E}^{(0)}_{\text{tot}}(a)}{A}=-\frac{(1-2ap^{0})}{12 \pi^{2}}\int^{\infty}_{0} \bar{k}^{3}d\bar{k}\int^{+\infty}_{-\infty}dz \\+\frac{(1-2ap^{0})}{12 \pi^{2}}\int^{\infty}_{0}\frac{\bar{k}^{2}d\bar{k}}{\Delta_{0}}\Bigg[\frac{\lambda_{1}}{2\bar{k}}+\frac{\lambda_{2}}{2\bar{k}}+2\frac{\lambda_{1}}{2\bar{k}}\frac{\lambda_{2}}{2\bar{k}}\bigg\{1-e^{-2\bar{k}L}\bigg\}-2\bar{k}L\frac{\lambda_{1}}{2\bar{k}}\frac{\lambda_{2}}{2\bar{k}}e^{-2\bar{k}L}\Bigg]. \label{Chap-4-A20}
\end{multline}
We observe that the first term in the above equation is a divergent quantity, which we define as bulk energy per unit area $\Big(\frac{\hat{E}^{(0)}_{\text{bulk}}}{A}\Big)$ associated with the parallel plates. After some calculational simplification, we find the $\kappa$-deformed total energy per unit area associated with the two parallel plates as
\be
 \frac{\hat{E}^{(0)}_{\text{tot}}}{A}=\frac{\hat{E}^{(0)}_{\text{bulk}}}{A}+\frac{\hat{E}^{(0)}_{\text{d1}}}{A}+\frac{\hat{E}^{(0)}_{\text{d2}}}{A}+\frac{\hat{E}^{(0)}_{C}}{A}, \label{Chap-4-A21}
\ee
where 
\be
\frac{\hat{E}_{\text{di}}}{A} = \frac{(1-2ap^{0})}{96\pi^{2} L^{3}}\int^{\infty}_{0}dY Y^{2}\frac{1}{\Big(1+\frac{Y}{\lambda_{i}L}\Big)},~~i=1,2\label{Chap-4-A22}
\ee
and 
\be
\frac{\hat{E}_{C}}{A} = -(1-2ap^{0})\frac{1}{96\pi^{2} L^{3}}\int^{\infty}_{0}dY Y^{3}\frac{1+\frac{1}{Y+\lambda_{1}L} +\frac{1}{Y+\lambda_{2}L}}{\Big(1+\frac{Y}{\lambda_{1}L}\Big)\Big(1+\frac{Y}{\lambda_{2}L}\Big)e^{Y}-1},\label{Chap-4-A23}
\ee
with $Y=2\bar{k}L$. Here $\hat{E}^{(0)}_{\text{d1}}$, $ \hat{E}^{(0)}_{\text{d2}}$ are the self-energies of the single plates, and $\hat{E}^{(0)}_{C}$ is the Casimir energy per unit area between the two parallel plates. Note that we have included terms up to the first order in deformation parameter $a$. In the strong interaction limit, i.e., $\lambda_{1},\lambda_{2} \rightarrow \infty$, from the expression of $\hat{E}^{(0)}_{C}$ we find the Casimir energy for the two parallel plates in $\kappa$-space-time. Further in the limit $a \rightarrow 0$, this $\kappa$-deformed casimir energy reduce to the commutative result \cite{CH-4-Mil}.

We substitute the physically relavant parts $\frac{\hat{E}_{\text{d1}}}{A}$, $\frac{\hat{E}_{\text{d2}}}{A}$ and $\frac{\hat{E}_{C}}{A}$ of Eq.(\ref{Chap-4-A21}) in Eq.(\ref{Chap-4-Ft00}) and find the $\kappa$-deformed centripetal force per unit area on the parallel plates as
\bea
\hat{\bar{\textbf{F}}}(\bar{t}) &=& - \hat{\bar{\textbf{r}}}(\omega\bar{t}) \, \omega^2 \bar{r}_{cm}\left[\hat{E}_{d1}^{(0)} + \hat{E}_{d2}^{(0)} + \hat{E}_{C}^{(0)} \right]\nonumber \\
&=& -(1-2ap^{0}) \, \omega^2 \bar{r}_{cm}\left[E_{d1}^{(0)} + E_{d2}^{(0)} + E_{C}^{(0)} \right]\hat{\bar{\textbf{r}}}(\omega\bar{t}),
\label{Chap-4-F-para}
\eea
where $\bar{r}_{\rm cm}=r_0 + z_{\rm cm}$. Here $E_{\text{d1}}^{(0)}$, $E_{\text{d2}}^{(0)}$ are energies associated with the two parallel plates and $E_{C}^{(0)}$ is Casimir energy between the two parallel plates in commutative space-time \cite{CH-4-shajesh2b,CH-4-shajesh3}. We observe that the non-commutative corrections to the force appear through an overall multiplicative factor $(1-2ap^{0})$. We also find the center of inertia in the curvilinear coordinates, $z_{\rm cm}$ for parallel plates. For this using Eq.(\ref{Chap-4-0thA11}) to Eq.(\ref{Chap-4-0thA13}) first we obtain
\be
\int^{+\infty}_{-\infty} z\,g^{(0)}(z,z,\bar{k})= \frac{L}{\Delta_{0}}\,\frac{1}{4\bar{k}^{2}}\,\bigg(\frac{\lambda_{1}}{2 \bar{k}}-\frac{\lambda_{2}}{2\bar{k}}\bigg)\label{Chap-4-inert}
\ee
Using Eq.(\ref{Chap-4-0thA10}) with above equation in Eq.(\ref{Chap-4-E-Q}) we find the center of inertia in the $\kappa$-deformed curvilinear coordinates as
\be
z_{\rm cm} \, \hat{E}_{tot}^{(0)} = 
- \frac{(1-2ap^{0})}{24\pi^2} \int_0^\infty \bar{k}^2 d\bar{k} \frac{L}{\Delta_{0}}
\left[ \frac{\lambda_2}{2\bar{k}} - \frac{\lambda_1}{2\bar{k}} \right].
\label{Chap-4-zcm-para}
\ee
In the above, note in general $\lambda_1\ne\lambda_2\ne 0$. But note that $z_{cm}$ will reduce to zero for $\lambda_1 = \lambda_2$, or in the Dirichlet limit. Using Eq.(\ref{Chap-4-A20}) in the above equation, we find that the $a$-dependent corrections on both sides get canceled. Thus we find that $z_{\rm cm} \, \hat{E}_{tot}^{(0)}$ does not get modified due to non-commutativity of the space-time.

Note that in Eq.(\ref{Chap-4-A23}) Casimir energy is attractive and the non-commutative contribution reduces the commutative value of the Casimir energy. We observe that the correction terms, which are coming due to the non-commutativity of space-time, reduce the centripetal force acting on the two parallel plates. It is important to note that as in commutative case \cite{CH-4-shajesh3} here also the $\kappa$-deformed Casimir energy including divergent parts (self energies of two plates) experiences the centripetal force exactly like a conventional mass in accordance with the mass-energy equivalence principle. In all the above calculations we included correction terms valid up to the first order in the deformation parameter $a$. The above expression of centripetal force acting on the Casimir apparatus will reduce to the commutative result \cite{CH-4-shajesh3} in the limit $a \rightarrow 0$.

\section{$\kappa$-defomed Centripetal force on oriented Casimir apparatus} \label{oricentforce}

\begin{figure}[h!]
\begin{center}
\includegraphics[scale=.7]{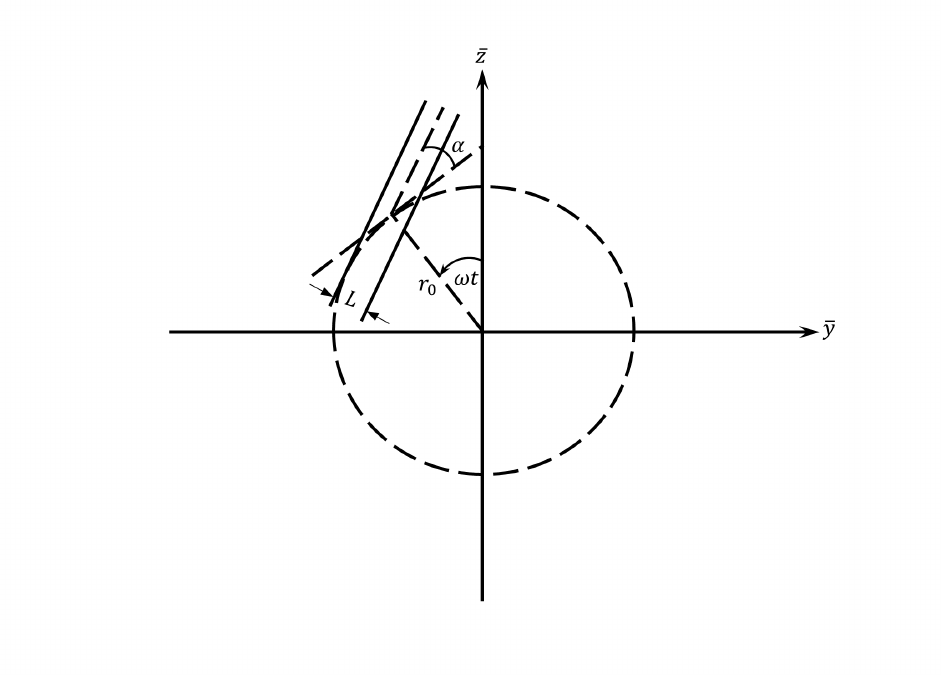}
\caption{\label{ori.rot-c}
A Casimir apparatus oriented at an angle $\alpha$ with respect to 
the tangent to the circle (with radius $r_0$) of motion and rotating with 
constant angular speed $\omega$ about the $\bar{x}$ axis.}
\end{center}
\end{figure}

In this Section, we derive the centripetal force acting on the parallel plates, which are tilted by an arbitrary angle. We consider that the Casimir apparatus is oriented at an arbitrary angle $\alpha$ with respect to the tangent to the circle (having a radius of $r_{0}$) of rotation while rotating with constant angular speed $\omega$ about the $\bar{x}$ axis as shown in the Fig-\ref{ori.rot-c}. 
Thus, by following the same procedure discussed in Section \ref{centforce}, we find the action for the rotation of the Casimir apparatus with orientation in $\kappa$-deformed local Lorentz coordinates as
\be
 \mathcal{S}_{\text{$\kappa$-tilt-LLC}}=\int d^4\bar{x} \sqrt{-\hat{\eta}(a)}
\left[
- \frac{1}{2} \hat{\eta}^{\alpha \beta}(a)\, 
\bar{\partial}_{\alpha} \bar{\phi}(\bar{x}) \bar{\partial}_{\beta} \bar{\phi}(\bar{x})
- \frac{1}{2} \bar{V}(\bar{x})\, \bar{\phi}(\bar{x})^2
\right], \label{Chap-4-sec7-L9}
\ee
where $\hat{\eta}^{\mu\nu} = \eta^{\mu \nu}+2ap^{0}(0,1,1,1)$ obtained from Eq.(\ref{Chap-4-N13}) and $\sqrt{-\hat{\eta}(a)}=(1-3ap^{0})$ valid upto first order in deformation parameter $a$. Here the background potential ($\bar{V}(\bar{x})$) is described as 
\bea
\bar{V}(\bar{x})&=& \lambda_1\,\delta 
(\bar{z}\cos(\alpha+\omega\bar{t}) 
 - \bar{y}\sin(\alpha+\omega\bar{t}) - r_1 \cos\alpha)\nonumber \\&&+\lambda_2\,\delta 
(\bar{z}\cos(\alpha+\omega\bar{t}) 
 - \bar{y}\sin(\alpha+\omega\bar{t}) - r_2 \cos\alpha)
\nonumber \\
&=& \lambda_1\,\delta
\left(\bar{z}\cos(\alpha+\omega\bar{t})
 - \bar{y}\sin(\alpha+\omega\bar{t}) - r_0 \cos\alpha + \frac{L}{2} \right)\nonumber \\&&+ \lambda_2\,\delta
\left(\bar{z}\cos(\alpha+\omega\bar{t})
 - \bar{y}\sin(\alpha+\omega\bar{t}) - r_0 \cos\alpha + \frac{L}{2} \right),
\label{Chap-4-V-bar-or}
\eea
where $r_{0}=\frac{r_{1}+r_{2}}{2}$ and $L=(r_{2}-r_{1})\cos \alpha$. $\alpha$ and $\omega$ are indicated in the Fig-\ref{ori.rot-c}. Next we make the following transformations between the local Lorentz coordinates and curvilinear coordinates
\begin{eqnarray}
y &= +\bar{y}\cos(\alpha+\omega\bar{t})
     +\bar{z}\sin(\alpha+\omega\bar{t})
     - r_0 \sin\alpha, \qquad x&=\bar{x},
\nonumber \\
z &= +\bar{z}\cos(\alpha+\omega\bar{t})-\bar{y}\sin(\alpha+\omega\bar{t})
     - r_0 \cos\alpha, \qquad t&= \bar{t},
\end{eqnarray}
which has the inverse transformations
\begin{eqnarray}
\bar{y} &= (y+r_0\sin\alpha) \cos(\alpha+\omega\bar{t})
           -(z+r_0\cos\alpha) \sin(\alpha+\omega\bar{t}),
           \qquad \bar{x}&=x, \nonumber
\\
\bar{z} &= (y+r_0\sin\alpha) \sin(\alpha+\omega\bar{t})
           +(z+r_0\cos\alpha) \cos(\alpha+\omega\bar{t}),
           \qquad \bar{t}&=t.
\end{eqnarray}
Note that for $\alpha =0$, the above transformations reduce to the transformations given in Eq.(\ref{Chap-4-trans-b}) and Eq.(\ref{Chap-4-in-trans-b}), respectively. Thus using these transformations in Eq.(\ref{Chap-4-sec7-L9}), we find the action in the $\kappa$-deformed curvilinear coordinates as 
\be
S_{\text{$\kappa$-tilt-Curvilinear}} = \int d^4x \sqrt{-\hat{g}(x,a)} \, {\cal{L}}(\phi(x),\partial_\mu \phi(x),a),\label{Chap-4-sec7-L10}
\ee
where the Lagrangian is
\be
{\cal{L}} 
= - \frac{1}{2} \hat{g}^{\mu\nu}(x,a) \partial_\mu \phi(x) \partial_\nu \phi(x)
- \frac{1}{2} V(x) \phi(x)^2, 
\label{Chap-4-sec7-L}
\ee
and the potential in curvilinear coordinates is
\begin{equation}
V(x) = \lambda_1\,\delta \left(z + \frac{L}{2}\right) 
       + \lambda_2\,\delta \left(z - \frac{L}{2} \right),
\label{Chap-4-sec7-pot}
\end{equation}
The $\kappa$-deformed metric  in curvilinear coordinates in the Eq.(\ref{Chap-4-sec7-L}) is
\be
\small
{\hat{g}_{\mu\nu}(x^{\alpha},a) = 
\left[ \begin{array}{cccc}
-\{1-(1-2ap^{0})\omega^2r^2\} &0& -(1-2ap^{0})\omega (z+r_0\cos \alpha) & (1-2ap^{0})\omega (y+r_{0}\sin \alpha )\\ 0&(1-2ap^{0})&0&0 \\
-(1-2ap^{0})\omega (z+r_0\cos \alpha) &0& (1-2ap^{0}) & 0 \\ (1-2ap^{0})\omega (y+r_{0}\sin \alpha) &0& 0 & (1-2ap^{0})
\end{array} \right],}
\label{Chap-4-rot-sec7-met}
\ee
where $r^2 = (y+r_0\sin\alpha)^2 + (z+r_0\cos\alpha)^2$ and $\sqrt{-\hat{g}(x,a)}=(1-3ap^{0})$. Here also by varying the action given in Eq.(\ref{Chap-4-sec7-L10}) with respect of the background metric given in Eq.(\ref{Chap-4-rot-sec7-met}) we find  energy-momentum tensor as
\be
\hat{t}_{\mu \nu}(x,a)
= \partial_{\mu} \phi(x) \partial_{\nu} \phi(x)
  + \hat{g}_{\mu \nu}(x,a) {\cal{L}}(\phi(x)),
\label{Chap-4-sec7-L13} 
\ee
where $\mathcal{L}$ is given in Eq.(\ref{Chap-4-sec7-L}). Next by repeating the procedure discussed in Section \ref{centforce}, we calculate the centripetal force as
\begin{equation}
\hat{\bar{\textbf{F}}}(\bar{t})
= - \omega^2 \, \Big(r_0 \hat{\bar{\textbf{r}}}(\omega \bar{t})
+ z_{ \text{cm}} \hat{\bar{\textbf{r}}}(\alpha + \omega \bar{t})\Big)
\Big[\hat{E}_{\text{d1}}^{(0)} + \hat{E}_{\text{d2}}^{(0)} + \hat{E}_{C}^{(0)} \Big],
\label{Chap-4-0r-Ft00}
\end{equation}
Note here that the $\kappa$-deformed centripetal force acting on the Casimir apparatus depends on the orientation angle $\alpha$. $\hat{E}_{\text{d1}}^{(0)}$, $\hat{E}_{\text{d2}}^{(0)}$ and $\hat{E}_{C}^{(0)}$  appearing in the above equation are given in Eq.(\ref{Chap-4-A22}) and Eq.(\ref{Chap-4-A23}), respectively. The center of inertia in the curvilinear coordinates, $z_{\text{cm}}$, is given in Eq.(\ref{Chap-4-zcm-para}). For $\alpha=0$, the force expression in the above equation reduces to the result obtained in Eq.(\ref{Chap-4-F-para}). We observe that for $z_{\text{cm}}=0$ the $\kappa$-deformed centripetal force is independent of the orientation of the Casimir apparatus. In all the above calculations, we have considered modifications valid up to the first order in deformation parameter $a$. The above force expression matches with the result obtained in \cite{CH-4-shajesh3} in the limit $a \rightarrow 0$.

\section{Conclusions} \label{conclu}

In this study, we investigated the effects of the non-commutativity of space-time on the centripetal force experienced by a rotating Casimir apparatus. We considered the Casimir apparatus to be rotating at a constant angular speed with respect to one of the space axes. We modeled this setup using $\kappa$-deformed local Lorentz and curvilinear coordinates. We constructed the force expression measured in the $\kappa$-deformed local Lorentz coordinates. Then using the transformations between local Lorentz coordinates and curvilinear coordinates we obtained the $\kappa$-deformed centripetal force on the Casimir apparatus. 

We calculate the centripetal force experienced by the plates (single and double plate configurations) in the presence of $\kappa$-deformed Klein-Gordon field. All our calculations are valid up to the first order in the deformation parameter $a$. First, we construct $\kappa$-deformed Klein-Gordon Lagrangian describing the Casimir apparatus rotating with constant angular speed. Using this Lagrangian we derive the energy-momentum tensor in $\kappa$-deformed curvilinear coordinates. We considered that the angular speed of the rotation is very small. Then we derive the $\kappa$-deformed Green's function, valid up to the first order in the deformation parameter, corresponding to the $\kappa$-deformed Klein-Gordon field. We use Green's functions obtained in different regions of the plates in calculating the vacuum expectation value of the energy-momentum tensor, which we use to find the centripetal force experienced by the plates (single and double plate configurations) as measured in the local Lorentz coordinates. We show that the $\kappa$-deformed centripetal force on the Casimir apparatus is the sum of the force acting on deformed Casimir energy and the force acting on the self-energies of the plates. We have also calculated the $\kappa$-deformed centripetal force on the oriented parallel plates and find that the centripetal force on the parallel plates depends on the orientation angel $\alpha$.

We note that even in non-commutative space-time, the Casimir energy, including the divergent parts, experiences the centripetal force like a conventional mass, as in the commutative case \cite{CH-4-shajesh3}. This shows that the equivalence principle holds in $\kappa$-defromed space-time also. Compared to the Casimir energy in the commutative space-time, we observe that the correction to the Casimir energy due to the non-commutativity in Eq.(\ref{Chap-4-A23}) has the opposite sign. As a result, the magnitude of Casimir energy in non-commutative space-time is lower than it is in commutative space-time \cite{CH-4-shajesh3}. We also observe that the product of the center of inertia and the total energy, that is,  $z_{\rm cm} \, \hat{E}_{tot}^{(0)}$ is not modified due to the non-commutativity of the space-time. It is significant to note that modifications in the centripetal force expression given in Eq.(\ref{Chap-4-F-para}) depend on $ap^{0}$ with the negative sign. Because of this, the net downward force on the parallel plates in the $\kappa$-deformed space-time is less than that in the commutative space-time. Here, $p^{0}$ is the energy of the probe that sees the non-commutative structure of the space-time and $a$ is the fundamental length scale in $\kappa$-space-time. In the modifications, the factor $ap^{0}$ determines how much of the centripetal force on the Casimir apparatus is reduced. Consequently, for $a \neq 0$, the energy of the probe determines amount of decrease in the centripetal force on the Casimir apparatus. The $\kappa$-deformed centripetal force on the Casimir apparatus given in Eq.(\ref{Chap-4-Ft00}) reduces to the results obtained in \cite{CH-4-shajesh3} in the commutative limit, that is, $a \rightarrow 0$.

\subsection*{\bf Acknowledgments}
S.K.P thanks UGC, India, for the support through the JRF scheme (id.191620059604).

\renewcommand{\thesection}{Appendix-A}
\section{Deformation of metric in $\kappa$-space-time}\label{append-A}
\renewcommand{\thesection}{A}

In this appendix, using the generalized commutation relation between the $\kappa$-deformed phase-space coordinates \cite{CH-4-zuhair1}, we construct the generic deformed metric in the $\kappa$-deformed space-time. The coordinates of $\kappa$-deformed space-time satisfy the commutation relations given in Eq.(\ref{Chap-4-intro2}). The generalised commutation relation for the $\kappa$-deformed phase space coordinates is \cite{CH-4-hopf,CH-4-Kova,CH-4-zuhair1,CH-4-kappa-geod},
\begin{equation}\label{Chap-4-N1}
 [\hat{\bar{x}}_{\mu},\hat{P}_{\nu}]=i\hat{g}_{\mu\nu}(\hat{\bar{x}}),
\end{equation} 
where $\hat{g}_{\mu\nu}$ is the $\kappa$-deformed metric and it is a function of the $\kappa$-deformed space-time coordinate $\hat{\bar{x}}_{\mu}$. 
We choose a specific realisation for the $\kappa$-deformed phase-space coordinates as \cite{CH-4-zuhair1},
\begin{equation}\label{Chap-4-N2}
 \hat{\bar{x}}_{\mu}=\bar{x}_{\alpha}\varphi^{\alpha}_{\mu},~ \,\hat{P}_{\mu}=g_{\alpha\beta}(\hat{\bar{y}})p^{\beta}\varphi^{\alpha}_{\mu},
\end{equation}
where $\hat{P}_{\mu}$ is the $\kappa$-deformed generalised momenta and $p_{\mu}$ is the canonical conjugate momenta corresponding to the commutative coordinate $\bar{x}_{\mu}$. In the commutative limit, i.e., $a\to 0$, we obtain $\hat{\bar{x}}_{\mu}\to \bar{x}_{\mu}$ and $\hat{P}_{\mu}\to p_{\mu}$. Substituting Eq.(\ref{Chap-4-N2}) in the $\kappa$-deformed space-time commutation relations, i.e., in Eq.(\ref{Chap-4-intro2}) we obtain $\varphi_{\mu}^{\alpha}$ as
\begin{equation}\label{Chap-4-N3}
 \varphi _0^0=1, \, \varphi _i^0=0, \, \varphi_0^i=0, \, \varphi _j^i=\delta _j^i e^{-\frac{ap^0}{2}}. 
\end{equation}
Note  that in Eq.(\ref{Chap-4-N2}), we introduced another set of $\kappa$-deformed space-time coordinates, $\hat{\bar{y}}_{\mu}$. It is assumed that the coordinates $\hat{\bar{y}}_{\mu}$ satisfy the $\kappa$-deformed space-time commutation relation as $[\hat{\bar{y}}_0,\hat{\bar{y}}_i]=ia\hat{\bar{y}}_i,~[\hat{\bar{y}}_i,\hat{\bar{y}}_j]=0$. It is also assumed that $\hat{\bar{y}}_{\mu}$ commutes with $\hat{\bar{x}}_{\mu}$, that is, $[\hat{\bar{y}}_{\mu},\hat{\bar{x}}_{\nu}]=0$. These new coordinates are introduced only to simplify the calculation\cite{CH-4-zuhair1}. The $g_{\alpha\beta}(\hat{\bar{y}})$ appearing in Eq.(\ref{Chap-4-N2}) has the same functional form as the metric in the commutative coordinate, but $\bar{x}_{\mu}$ replaced with non-commutative coordinate $\hat{\bar{y}}_{\mu}$. We now express $\hat{\bar{y}}_{\mu}$ in terms of the commutative coordinate and its conjugate momenta as
\begin{equation}\label{Chap-4-N4}
 \hat{\bar{y}}_{\mu}=\bar{x}_{\alpha}\phi_{\mu}^{\alpha}.
 \end{equation}
Using $[\hat{\bar{y}}_0,\hat{\bar{y}}_i]=ia\hat{\bar{y}}_i,~[\hat{\bar{y}}_i,\hat{\bar{y}}_j]=0$ and $[\hat{\bar{x}}_{\mu},\hat{\bar{y}}_{\nu}]=0$, one obtains $\phi_{\mu}^{\alpha}$ as  \cite{CH-4-zuhair1}

\begin{equation}\label{Chap-4-N5}
 \phi_{0}^{0}=1,~\phi_{i}^{0}=0,~\phi_{0}^{i}=-\frac{a}{2}p^i,~\phi_{i}^{j}=\delta_{i}^{j}.
\end{equation}
Thus, we find the explicit form of $\hat{\bar{y}}_{\mu}$ as
\begin{equation}\label{Chap-4-N6}
 \hat{\bar{y}}_0=\bar{x}_0-\frac{a}{2}\bar{x}_jp^j,~~
\hat{\bar{y}}_i=\bar{x}_i.
 \end{equation}
Using the above in Eq.(\ref{Chap-4-N2}) and substituting $\hat{\bar{x}}_{\mu}$ and $\hat{P}_{\mu}$ in Eq.(\ref{Chap-4-N1}), we find the $\kappa$-deformed metric \cite{CH-4-zuhair1}
\begin{equation}\label{Chap-4-N7}
 [\hat{\bar{x}}_{\mu},\hat{P}_{\nu}] \equiv i\hat{g}_{\mu\nu}=ig_{\alpha\beta}(\hat{\bar{y}})\Big(p^{\beta}\frac{\partial \varphi^{\alpha}_{\nu}}{\partial p^{\sigma}}\varphi_{\mu}^{\sigma}+\varphi_{\mu}^{\alpha}\varphi_{\nu}^{\beta}\Big). \end{equation}
Note that $g_{\mu\nu}(\hat{\bar{y}})$ can be obtained by replacing the commutative coordinates with the $\kappa$-deformed coordinates for any given commutative metric.
Using Eq.(\ref{Chap-4-N3}) in Eq.(\ref{Chap-4-N7}) we find the components of $\hat{g}_{\mu\nu}$ as
\begin{equation}\label{Chap-4-N9}
\begin{aligned}
\hat{g}_{00}&=g_{00}(\hat{\bar{y}}),\\
\hat{g}_{0i}&=g_{0i}(\hat{\bar{y}})\big(1-ap^0\big)-\frac{a}{2}g_{im}(\hat{\bar{y}})p^{m},\\ 
\hat{g}_{i0}&=g_{i0}(\hat{\bar{y}})\big(1-\frac{ap^0}{2}\big),\\
\hat{g}_{ij}&=g_{ij}(\hat{\bar{y}})\big(1-ap^0\big).
\end{aligned}
\end{equation}
Note here that we consider terms valid up to the first order in $a$ only.

\end{document}